\documentstyle[12pt]{article}

\newcommand{\be}{\begin{equation}}
\newcommand{\ee}{\end{equation}}
\newcommand{\ba}{\begin{eqnarray}}
\newcommand{\ea}{\end{eqnarray}}
\begin{document}
\def\pct#1{\input epsf \centerline{ \epsfbox{#1.eps}}}

\begin{titlepage}
\hbox{\hskip 12cm ROM2F-97/07  \hfil}
\hbox{\hskip 12cm \today \hfil}
\vskip 1.5cm
\begin{center} 
{\Large  \bf  Geometrical \ Construction \ of \ Type I \\
~ \\
Superstring \ Vacua \footnote{Talk
presented at the ``12th Italian Conference on General Relativity and
Gravitational Physics '', Rome, september 23-27,
1996.}}
 
\vspace{1.8cm}
 
{\large \large Gianfranco \ \  Pradisi}

\vspace{0.8cm}

{\sl Dipartimento di Fisica\\
Universit{\`a} di Roma \ ``Tor Vergata'' \\
I.N.F.N.\ - \ Sezione di Roma \ ``Tor Vergata'' \\
Via della Ricerca Scientifica, 1 \ \
00133 \ Roma \ \ ITALY}
\vspace{0.5cm}
\end{center}
\vspace{3cm}
\abstract{The parameter-space orbifold construction of open and unoriented
toroidal and (target-space) orbifold compactifications is briefly reviewed, with
emphasis on the underlying geometrical framework.  A class of 
chiral four-dimensional type-I vacua with three generations
is also discussed.} 
\vfill \end{titlepage}  
\hbox{\hskip 12cm ROM2F-97-07  \hfil}
\begin{center}

{\large  \bf   Geometrical \ Construction \ of \ Type I \\
Superstring \ Vacua}
\vspace{0.5cm}

{\large  Gianfranco \ \  Pradisi}
\vspace{0.3cm}

{\sl Dipartimento di Fisica, \ \ Universit{\`a} di Roma \ ``Tor Vergata''
\\ I.N.F.N.\ - \ Sezione di Roma \ ``Tor Vergata'', \ \ Via della Ricerca
Scientifica, 1
\\ 00133 \ Roma, \ \ ITALY}
\vspace{0.5cm}
\end{center}
\begin{abstract}
{\small The parameter-space orbifold construction of open and unoriented
toroidal and (target-space) orbifold compactifications is briefly reviewed, with
emphasis on the underlying geometrical framework.  A  class of 
chiral four-dimensional type-I vacua with three generations
is also discussed.} 
\end{abstract} 

\addtolength{\baselineskip}{0.3\baselineskip} 

Superstring theories \cite{gsw} are 
defined perturbatively as
Conformal Field Theories with a field content that saturates the conformal
anomaly.  Superstring amplitudes are correlation functions of
suitable combinations of chiral vertex operators on arbitrary Riemann
surfaces \cite{fms}.  Limiting the perturbative expansion to only closed
oriented surfaces correspond to defining closed oriented (type II 
or heterotic) superstring models.  Whenever the
bulk algebra exhibits left-right symmetry, it is possible to extend the
expansion to non-orientable and bordered Riemann surfaces, thus describing also
unoriented closed and open (type I) superstrings. 
The consistency conditions can be fully encoded in a set of {\it sewing
constraints} connecting the `basic building blocks', namely  coefficients
of suitable correlation functions on low genus surfaces, both for closed
oriented models \cite{sonoda} and for unoriented closed and open models
\cite{fps} \cite{pss2} \cite {pss1} \cite{lew}.  
Equivalently, the interest can be turned
to the analysis of perturbative spectra, encoded in the one-loop partition
function.  This approach deserves several observations: first, closed
superstrings are characterized by modular invariant partition functions
\cite{gso}, a property no longer true for amplitudes on surfaces endowed with 
holes and crosscaps. 
Second, it has long been known that models of only open superstrings are
non-unitary, due to the presence of closed superstring states in
intermediate channels of open diagrams \cite{lovel}.  Third,
multiplicities associated to boundary fields (open vertex operators) are
promoted to gauge (Chan Paton) factors of some classical Cartan groups
\cite{cp}.  These are (partly) fixed by tadpole conditions, that enforce the
inclusion of open and unoriented contributions{\footnote{The absence of tadpole
conditions allows in fact type I models with only closed unoriented
superstrings \cite{noopen} \cite{gep}.}.  As a result, type I vacua are
{\it parameter space orbifolds} \cite{cargese} of corresponding `parent'
type IIB vacua.  Four contributions enter their one-loop partition
function.  The first one is the torus amplitude, that encodes the spectrum of
the closed oriented `parent' model.  In order to construct a class of
`open descendants', one projects the closed spectrum into an
unoriented one adding to the (halved) torus the 
Klein bottle amplitude.  Then, the two open
contributions, annulus and M\"obius strip amplitudes, complete the
description of the open unoriented spectrum.  The construction resembles
closely what happens in $Z_2$ orbifolds \cite{dhvw}, where the closed
spectrum is projected in a $Z_2$-invariant way and `twisted' sectors
corresponding to strings closed only on the orbifold are added and
projected.  The $Z_2$ group is enforced by the `twist operator'
$\Omega$ (now commonly referred to as world-sheet parity operator) that 
interchanges 
left and right sectors , while open superstrings, closed only on the
double cover, play the role of `twisted sectors'.  Thus, the orbifold
should be thought of in parameter space rather than in target space,
and the natural action of `twist' requires that the closed `parent' model 
be left-right symmetric.  It is worth noticing that all open descendants of 
non left-right symmetric `parent' models discussed in the literature 
are related by T-duality \cite{tdual} to left-right 
symmetric ones.

Type I vacua were considered,
in the past, less appealing than heterotic ones,
essentially for phenomenological reasons.  In the last two years,
however, much progress has been made in understanding the
non-perturbative structure of superstrings.  All superstring
vacua emerge as asymptotic 
expansions around specific points of moduli space of
an unknown fundamental (M \cite{mtheory} or F \cite{ftheory}) theory. 
Moreover, a number of duality symmetries \cite{dual} connecting the various
perturbative vacua have been found or conjectured that make plausible such a
unified scenario, and that correspond to different compactifications of M or F
theory.  For instance, there is a strong-weak coupling duality
between the type I $SO(32)$ superstring and the $SO(32)$ heterotic string
in $d=10$ \cite{wittendy}, and a T-duality 
connecting the latter to the $E_8 \times E_8$
heterotic string in $d=9$ \cite{nsw}.  
All three can be thought of as `derived' from
the compactification of eleven dimensional M theory on 
$S^{1}/Z_{2}$, that gives rise to the $E_8
\times E_8$ heterotic string \cite{horwit}.  

In this article we shall review some aspects of the 
parameter space orbifold construction.  Our aim is
introducing the relevant ideas to describe a class of
chiral four dimensional type I vacua recently presented in ref.
\cite{chiral}.  Let us begin with the simplest irrational case, namely the
class of open descendants of the toroidal compactification on a
one-dimensional circle of radius $R$ \cite{toroidal}.  The (halved) 
closed partition 
function for the internal part is given by  
\be
{\cal{T}} \ = \ {1 \over 2} \ 
{{\sum Êq^{p_{L}^{2} / 2} \ {\bar{q}}^{p_{R}^{2} / 2} \over {\eta( \tau )
\eta( \bar{\tau} )}}} \qquad ,  
\label{torusc1}
\ee
where $p_{L,R} \: = \: ( \, \sqrt{\alpha^{'} / 2} \, ) \: 
( \, m/R \, \pm \, nR/{\alpha^{'}} \, )$.  The
action of $\Omega$ on the Hilbert space of states
consists of interchanging  holomorphic and antiholomorphic parts.  
The `crosscap 
constraint' of refs. \cite{fps,pss1} allows some freedom in the 
choice of $\Omega$ eigenvalues.  We limit ourselves to the
simplest `standard' Klein bottle projection with all sectors
symmetrized, while bearing in mind that in general 
there exist several choices for the non-oriented
projection.  
The Klein bottle contribution 
\be 
{\cal K} \ = \ {1 \over 2} \ \
{{\sum\limits_{m}Êe^{-  \pi \tau \alpha^{'} \, ({m \over R})^2}}
\over {\eta( 2 i \tau )}}
\label{kleinc1}
\ee
completes the spectrum of the unoriented closed sector.  While the
torus amplitude is modular invariant, an $S$-transformation turns the Klein
partition function into the closed crosscap-to-crosscap amplitude
\be
{\tilde{\cal	K }} \ = \ {v \over 2} \ 
{\sum\limits_{n} \ Êq^{ {( 2 n R )^2} \over {4 \alpha^{'}}}
\over {\eta( q )}} \qquad ,
\label{ktildec1}
\ee  
where $v$ denotes the (dimensionless) volume $R/\sqrt{\alpha^{'}}$. 
Notice that only the closed sectors labelled by even winding modes can
flow through the crosscap, as demanded by the freely acting involution.  The
presence in this channel of massless characters may signal some
inconsistency due to `tadpoles' of `unphysical' fields \cite{pcay}, the
cancellation of which demands the inclusion of open superstrings.  
For irrational values of $R$, the
closed spectrum is a superposition of an infinite number of sectors 
labelled by the winding mode and coupled with
their conjugates in the amplitude (\ref{torusc1}).  They can thus be
consistently reflected in front of a boundary or, which is the same, can flow
in the transverse annulus amplitude with (in general) individual reflection
coefficients.  This can be achieved by including continuous `Wilson Lines'
on the boundaries that shift the momentum according to the (constant) values
of background internal gauge fields.  To illustrate this mechanism, it
is useful to think of eq. (\ref{torusc1}) as the lattice sum of the Type IIB
superstring compactified on a circle.  We are thus omitting (as always in
this paper) all factors inert with respect to the modular properties. 
In
order to give rise to marginal deformations, the Wilson lines 
must be chosen in a Cartan
subalgebra of $SO(32)$, the unbroken gauge group of the ten-dimensional 
type I superstring \cite{gs}. 
A standard parametrization of
the Wilson line is, for each boundary, 
\be
U \ =  \  \ exp \bigoplus_{r=1,\cdots,16} i \ \alpha^r \ \sigma_2 \quad ,
\label{wline}
\ee
where $\alpha^r$ are proportional to the constant values $A^{r}$ of the gauge
field in the Cartan subalgebra.  
A contribution $Tr(U)^2$ (due to the presence of two boundaries) must be
included in the path integral for the annulus partition function and a
contribution $Tr(U^2)$ (due to the presence of a single boundary of double
length) must be included for the M\"obius strip.  
For instance, choosing $\alpha^r=0$ for
$r = 1 , \dots , N$, and $\alpha^r=\alpha$ for $r = N+1, \dots , 16$, 
the annulus
partition function reads 
\ba
{\cal{A}} \ &=& \ \Biggl[ \ \Big( \ {{(2N)^2} \over 2} \ + \
M \bar{M}\ \Bigr)  \ \sum\limits_{m}Êe^{-  \pi \tau \alpha^{'} \,
{m^2 \over R^2}}  \nonumber \\ &+& (2N) M  \ \sum\limits_{m}Êe^{-  \pi \tau
\alpha^{'} \, {{(m + a)^2} \over R^2}} +  (2N)
\bar{M} \ \sum\limits_{m}Êe^{-  \pi \tau
\alpha^{'} \, {{(m - a)^2} \over R^2}} \nonumber \\ &+&  {{M^2} \over
2} \
 \sum\limits_{m}Êe^{-  \pi \tau
\alpha^{'} \, {{(m + 2 a)^2} \over R^2}} + 
{{\bar{M}^2} \over 2} \
 \sum\limits_{m}Êe^{-  \pi \tau
\alpha^{'} \, {{(m - 2 a)^2} \over R^2}} \Biggr] \Big/ \eta\big ( {{i
\tau} \over 2}\big ) 	\quad , \label{annulusc1}
\ea
where $a = Q R A$, with $Q$ the charge and $A$ the internal gauge field.  
The M\"obius strip partition function completes the
projection in the non-oriented sector 
\ba
{\cal{M}} \ &=& \ - \ 
\Bigg[ \ {{2 N} \over 2} 
 \sum\limits_{m}Êe^{-  \pi \tau
\alpha^{'} \, {m^2 \over R^2}} 
+  
{{M} \over 2} \
 \sum\limits_{m}Êe^{-  \pi \tau
\alpha^{'} \, {{(m + 2 a)^2} \over R^2}} \nonumber \\ &+&
{{\bar{M}} \over 2} \
 \sum\limits_{m}Êe^{-  \pi \tau
\alpha^{'} \, {{(m - 2 a)^2} \over R^2}} \Bigg] \: \Big/ \, \hat{\eta}
\big( {{i \tau} \over 2} + {1 \over 2}\big) \quad , 
\label{mobiusc1}
\ea  
`hatted' terms being, as usual, the real basis for $\cal{M}$ \cite{bs}.  
From the boundary-to-boundary amplitude
\be
{\tilde{\cal A}} \ = \ {v \over 2} \ {\sum\limits_{n} \ \big[ 2N + M e^{2
\pi i a n} + \bar{M} e^{- 2 \pi i a n} {\big]}^2 \ \: {q^{{(n R)^2}
\over {4 \alpha^{'}}} \over {\eta(q)}}} \quad , 
\label{antrac1}
\ee
it is easy to read the tadpole condition for massless R-R sector.  Taking
into account the powers of two from the modular measure and the cooperative 
action of ${\tilde{\cal A}}$ with ${\tilde{\cal K}}$ and ${\tilde{\cal M}}$,
tadpole 
cancellation fixes the overall sign in eq. (\ref{mobiusc1}) and results in 
\be
2N \ + \ M \ + \ \bar{M} \ = \ 32  \qquad  .  
\label{tadpcon}
\ee
Wilson lines thus drive
the breaking of the Chan-Paton symmetry by moving part of the charge to 
massive sectors.  In this simple example, the unbroken gauge group is
generically $SO(2N) \otimes U(16-N)$.  Extra massless modes appear, however,
for specific values of the background gauge field, giving rise to a group
enhancement phenomenon.  For instance, in eqs. (\ref{annulusc1},
\ref{mobiusc1}) a half-integer value of $a$ leads to $SO(2N) \otimes
SO(32-2N)$, while an integer value of $a$ restores the unbroken $SO(32)$. 
It should be appreciated that the whole construction is manifestly
compatible with planar duality and factorization of amplitudes, as can be
tested analyzing the proper sewing constraints.  In the rational
case, the closed spectrum can be organized in terms of a finite number of
characters of an extended algebra.  There is a one to one correspondence
between the number of chiral sectors and the number of Chan-Paton charges,
that collapse to a corresponding 
finite number if the boundaries are to preserve the 
symmetry.  This is really a subtle
issue, because in some non-diagonal cases the algebra of open states is 
actually an extended algebra even if the bulk algebra is not \cite{pss2}.  

Open descendants can be built for more general toroidal compactifications
than those on (products of) circles \cite{toroidal}.  
A subtlety emerges due to
the presence of `torsion' in the lattice defining the torus, encoded in
the antisymmetric NS-NS tensor B.  In fact, the presence of B makes in
general the theory not left-right symmetric, a property recovered 
only for `quantized' values of the (constant) B field,
that ceases anyway to be a modulus.  Although projected out from the closed
non-oriented spectrum, the B field plays an important role in the open
sector, because its presence reduces the size of the Chan-Paton group
by a factor
$2^{r/2}$, with $r$ the rank of B.  This observation will be crucial in the
construction of chiral four dimensional models.

The geometrical nature of consistency conditions is more evident 
if we extend the parameter space orbifold construction to 
(target-space) irrational orbifolds.  While generalizations are
straightforward, we shall confine our attention to 
the one-dimensional $Z_2$ orbifold of
the circle of  radius $R$, following ref. \cite{ps}.  Several new 
ingredients enter the game.  First of all, in order to
understand the contributions in the open and unoriented sectors, it is
necessary to properly define the combined action of the (target-space)
orbifold group and the `twist'.  This can be achieved in the following
geometrical setting: Klein bottle, annulus and M\"obius strip are
$Z_2$-orbifolds of the double cover torus with respect to the action of an
anticonformal involution, the geometrical counterpart of $\Omega$.  Only
the orbifold sectors compatible with the involution can
`descend' on the corresponding surface.  They amount precisely to the 
sections of the $Z_2$ line bundle on that surface. 
This is exactly what other authors
recently called `gauging the orientifold group' \cite{gimpol}.  Let us see
explicitly the $Z_2$ case.  Defining the characters 
\ba \xi_{++} \
= \ {1 \over 2} \, ( {1 \over \eta } \ + \ \Theta_{34} ) \qquad , \qquad  
\xi_{-+} \ = \ {1 \over {2 \sqrt{2}}} \, ( \Theta_{23}  \ + \ \Theta_{24} )
\quad , \nonumber \\  \xi_{+-} \ = \ {1 \over 2} \, ( {1 \over \eta } \ - \
\Theta_{34} ) \qquad , \qquad \xi_{--} \ = \ {1 \over {2 \sqrt{2}}} \, (
\Theta_{23}  \ - \ \Theta_{24} )  \quad ,
\label{z2char} 
\ea
with $\Theta_{ij} = \theta_{i}^{1/2} \theta_{j}^{1/2} / \eta$ a combination
of standard elliptic functions, the Klein bottle partition function is
\be
{\cal K} \ = \ {1 \over 4 \eta} \ \big[ \ \sum\limits_{m} e^{- \pi \tau
\alpha^{'} ({m \over R})^2} \ + \  \sum\limits_{n} e^{- {{\pi
\tau}\over {\alpha^{'}}} (n R)^2} \ \big] \ + \ \xi_{-+} \ + \ \xi_{--}
\quad. \label{z2klein}
\ee
Notice that the first two terms are, respectively, 
the standard symmetrizations of states
with $p_L = p_R$ and the projections of states with $p_L=-p_R$, allowed only
on the orbifold.  The last two terms project the twisted (R-independent)
sectors where the fixed point ambiguity between spin fields has been resolved
in a diagonal fashion.  It should be appreciated that, as expected from the
experience with the rational case, 
in the tube terminating at two crosscaps only states with
even windings or momenta (thus from `untwisted sector') can flow, as encoded
in the transverse Klein bottle amplitude: 
\ba
{\tilde{\cal K}} \ &=& \ {1 \over 4 \eta} \ \big[ \ v \sum\limits_{n\ne 0}
q^{{1 \over {4 \alpha^{'}}} (2 n R)^2} \ + \ {1 \over v} \sum\limits_{m
\ne 0} q^{{\alpha^{'} \over 4}  ({2 m \over R})^2} \ \big] \nonumber \\
&+& \
{1 \over 4} \ ( \sqrt{v} + {1\over \sqrt{v}} )^2 \ \xi_{++} \ + \ {1 \over
4} \ ( \sqrt{v} - {1\over \sqrt{v}} )^2 \ \xi_{+-} \quad.  
\label{z2ktilde} 
\ea
Notice also that the coefficients in (\ref{z2ktilde}) are, as expected, 
squares of reflection 
coefficients in front of the crosscap.  
The sections in the open sector are very interesting because the
orbifold allows for open strings with generalized boundary conditions.  In
particular, three kinds of open strings are present, namely strings with
(standard) Neumann conditions at both ends (NN), strings with Dirichlet
conditions at both ends (DD) and strings with mixed conditions (ND). 
Recently open strings with Dirichlet conditions have received much attention
because their ends live on dynamical hyperplanes (Dirichlet p-branes or
D-branes \cite{dlp} \cite{dbrane}) whose excitations are open string
modes.  D-branes play also a fundamental role in connection 
with non-perturbative
aspects of superstring theories: they are in particular carriers of 
Ramond-Ramond charges \cite{polrr}.  There is another crucial ingredient in
the construction: the orbifold group acts in a non-trivial fashion on the
Chan-Paton charge space.  Indeed, in the simplest cases, the corresponding
charge assignment is driven by the fusion rules, but in general it is more
involved \cite{pss2}.  If we parametrize the multiplicities entering the
partition function for the `Neumann' and `Dirichlet' Chan-Paton groups
as   
\ba
&Tr({\bf 1}_{N}) \ = \ N_+ \ + \ N_- \quad, \qquad &Tr({\bf 1}_{D_{i}}) \ =
\ D_{i+} \ + \ D_{i-} \quad, \nonumber \\
&Tr({\bf R}_{N}) \ = \ N_+ \ - \ N_- \quad, \qquad &Tr({\bf R}_{D_{i}}) \ =
\ D_{i+} \ - \ D_{i-} \quad,
\label{traces}
\ea
where $i$ labels the two fixed points, the
annulus reads
\ba
{\cal A} \ &=& \ {1 \over 4 \eta} \ \big[ (N_+ + N_-)^2
\ \sum\limits_{m \ne 0} e^{- \pi \tau \alpha^{'} ({m \over R})^2}
\nonumber\\ &+& \sum\limits_{i,j} \, (D_{i+} + D_{i-}) \, (D_{j+} + D_{j-})
\ \sum\limits_{n \ne 0} e^{- {\pi \tau R^2 \over \alpha^{'}} (n +
\Delta_{ij})^2} \big] \nonumber\\ &+& [{N^{2}_{+} \over 2} + {N^{2}_{-}
\over 2} + \sum\limits_{i} \, ( {D_{i+}^2 \over 2} + {D_{i-}^2 \over 2})] \,
\xi_{++} \, + \, [N_{+} N_{-} + \sum\limits_{i} \, ( D_{i+} D_{i-} )] \,
\xi_{+-} \nonumber\\ &+& \, [\sum\limits_{i} (N_{+} D_{i+} + N_{-} D_{i-})]
\xi_{-+} \, + \, [\sum\limits_{i} (N_{+} D_{i-} + N_{-} D_{i+})] \xi_{--}
\quad. 
\label{z2annul} 
\ea
The $\Delta_{ij}$ vanish if both ends of the Dirichlet strings are at the same
fixed point, while they are 
$1/2$ if the ends are at different fixed points.  It is
simple to verify that the transverse amplitude is a linear superposition of an
(infinite) number of characters and the coefficients are perfect squares of
the reflection coefficients in front of the boundaries, as demanded by sewing
constraints.  Notice that only (NN) strings and (DD) strings with ends at 
the same fixed point can flow in the M\"obius strip.  The other types of open
strings are in fact oriented and do not contribute to the M\"obius partition
function 
\ba
{\cal M} \ &=& \ - {1 \over 4 \hat{\eta}} \ \big[ (N_+ + N_-)
\ \sum\limits_{m \ne 0} e^{- \pi \tau \alpha^{'} ({m \over R})^2}
\nonumber\\ &+& \sum\limits_{i} \, (D_{i+} + D_{i-}) 
\ \sum\limits_{n \ne 0} e^{- {\pi \tau R^2 \over \alpha^{'}} n^2} \big]
\nonumber\\ &-& [{N_{+} \over 2} + {N_{-} \over 2} + \sum\limits_{i}
\, ( {D_{i+} \over 2} + {D_{i-} \over 2})] \, \hat{\xi}_{++}  \quad. 
\label{z2mobius}
\ea
The setting just illustrated concerns real charges.  The involution,
however, allows in this case additional phases which turn the
Chan-Paton charges into complex ones.  Consistency of the vacuum channel is
ensured by the equality of the multiplicities of charges and anticharges
for each Chan-Paton factor.  Coming back to the real case in 
eqs. (\ref{z2annul}) and (\ref{z2mobius}), the
transverse amplitudes give rise to two tadpole conditions.  The first is
actually a pair of conditions, due to the incommensurability of the volume
and the inverse volume.  These conditions typically fix the
total dimensionality of the Chan-Paton charge space both for Neumann and
Dirichlet sector in a way analogous to eq. (\ref{tadpcon}).  The second
tadpole condition is actually present only when some twisted sector becomes
massless, and it is a constraint linking Neumann and Dirichlet groups.  

In order to use this construction in supersymmetric models, a
simultaneous twist of some `space-time' coordinates is needed in such a
way to preserve the antiperiodicity of the supercurrent.  Open
descendants of left-right symmetric superstring theories in arbitrary
dimensions can be constructed in this way.  In ref. \cite{bs} a
number of  models in six dimensions have been analyzed
starting from convenient rational points in the moduli space.  They exhibit
a rich structure of Chan-Paton gauge groups.  Introducing
suitable (discrete) open-string Wilson lines, it is possible to enhance the
Chan-Paton group up to $U(16) \otimes U(16)$.  The distinctive feature of
these type I models is the presence in the perturbative spectra of a variable
number of tensor multiplets.  They play a fundamental role in
the generalized Green-Schwarz anomaly cancellation 
mechanism \cite{ggs}.  Recently, using parameter-space orbifolds of Gepner
models, many other examples of type I vacua in six dimensions have been
constructed with variable numbers of tensor multiplets (including zero)  
and rich patterns of Chan-Paton symmetry breaking \cite{gep}. 
Several authors have also obtained six-dimensional type I models
using essentially the orbifold construction previously discussed
\cite{gimpol} \cite{noopen} \cite{6dmodels}.  Many of these models coincide
with those of ref. \cite{bs} if specialized to rational values of the moduli.  

Finally let us describe the {\it only} available class of
four dimensional type I chiral models, recently discovered by merging all
the ideas above \cite{chiral}.  We start from the $Z$-orbifold reduction 
of the type IIB
superstring.  In particular, the toroidal lattice comprises three
copies (of sizes $R_i$) of a two-dimensional hexagonal lattice, where $Z_3$ 
has a natural action.  Moreover, we choose a vanishing NS-NS
antisymmetric tensor in order 
to avoid small-sized Chan-Paton groups (recall that
the quantized values of $B$ reduce their maximum size by a factor
$2^{r/2}$).  The closed spectrum must exhibit $N=2$ supersymmetry, then
reduced to $N=1$ by the unoriented truncation.  To this end, as
anticipated, it is necessary to twist some internal world-sheet
fermions as well.  The (light-cone) $SO(8)$ characters must be decomposed
with respect to $SO(2)\otimes SU(3) \otimes U(1)$.  Introducing 
\ba
\Xi_{0,\epsilon}(q) &=& \left( {{ A_0 \chi_0 + \omega^{\epsilon} A_+ \chi_-
+  {\bar\omega}^{\epsilon} A_- \chi_+ } \over {{H_{0,\epsilon}}^3}} \right)
(q)  \nonumber \\ \Xi_{+,\epsilon}(q) &=& \left( {{ A_0 \chi_+ + 
\omega^{\epsilon} A_+ \chi_0 +  {\bar\omega}^{\epsilon} A_- \chi_- } \over
{{H_{+,\epsilon}}^3}} \right) (q)  \nonumber \\ \Xi_{-,\epsilon}(q) &=&
\left( {{ A_0 \chi_- + \omega^{\epsilon} A_-  \chi_0 + 
{\bar\omega}^{\epsilon} A_+ \chi_+} \over {{H_{-,\epsilon}}^3}} \right)(q) 
\quad ,
\label{csis}
\ea
where  $\{ A_0 , A_+ ,A_-\}$ are supersymmetric characters of conformal
weights $\{1/2,1/6,1/6 \}$ respectively, $\{\chi_0,\chi_{+},\chi_{-} \}$ are 
level-one SU(3) characters of conformal weights $\{0,1/3,1/3 \}$
respectively, and  
\ba
 H_{0,{\epsilon}}(q)  &=&  q^{1\over {12}}
\prod_{n=1}^{\infty}(1-\omega^{\epsilon} q^n)  (1 - {\bar \omega}^{\epsilon}
q^n)  \quad, \nonumber \\
H_{+,{\epsilon}}(q) &=& H_{-,{- \epsilon}}(q) = 
3^{-{1\over2}}{q^{-{1\over{36}}}}
\prod_{n=0}^{\infty}(1 - \omega^{\epsilon} q^{n+{1\over 3}}) (1 -  {\bar
\omega}^ {\epsilon} q^{n+{2\over3}}) \quad ,
\label{acca}
\ea
with $\epsilon = {0,\pm 1}$ and $\omega = e^{{2 \pi i}\over {3}}$, 
one can write the closed orbifold partition
function in the form 
\ba
T &=& {1 \over 3} \ \Xi_{0,0}(q) \ \Xi_{0,0}({\bar {q}}) \sum
q^{{{1} \over 2} p_{La} G^{ab} p_{Lb}}
{\bar{q}}^{{{1} \over 2} p_{Ra} G^{ab} p_{R b}}
 + \ {1 \over 3} \ \sum_{\epsilon = \pm 1} \ \Xi_{0,\epsilon}(q) \
\Xi_{0,\epsilon}({\bar{q}})
\nonumber \\  
& & \qquad + \ {1 \over 3} \ \sum_{\eta = \pm 1} \ \sum_{\epsilon =
0,\pm 1} \ 
\Xi_{\eta,\epsilon}(q) \ \Xi_{-\eta,-\epsilon}({\bar{q}}) \quad ,
\label{torochir}
\ea
where $G$ is the lattice metric.
Notice that the Klein-bottle projection contains only the sublattice with
$p_{L} = p_{R}$.  No other possibilities are allowed by the $Z_3$ involution. 
By defining for $\sigma = 0,\pm 1$ and $\eta = 0,\pm 1 \, ( \, mod \, 3 \, )$
\be
\rho_{\sigma , \eta} \ = \ \sum\limits_{\epsilon = 0, \pm 1} \ 
{{{\bar{\omega}}^{\eta\epsilon}} \over {H_{\sigma \epsilon}^{3}}}
\label{csisn}
\ee
we can introduce 
the following $R_i$-independent untwisted characters
\be
\Lambda_{0,\eta} \ = \ 
A_{0} \, \chi_{0} \, \rho_{0,\eta} \ + \ 
A_{+} \, \chi_{-} \, \rho_{0,\eta - 1} \ + \ 
A_{-} \, \chi_{+} \, \rho_{0,\eta + 1} \quad, 
\label{untchard4}
\ee
and the following (obviously $R_i$-independent) twisted characters
\ba
\Lambda_{+,\eta} \ &=& \ 
A_{0} \, \chi_{+} \, \rho_{+,-\eta} \ + \ 
A_{+} \, \chi_{0} \, \rho_{+,1-\eta} \ + \ 
A_{-} \, \chi_{-} \, \rho_{+,-1-\eta} \quad, \nonumber \\
\Lambda_{-,\eta} \ &=& \ 
A_{0} \, \chi_{-} \, \rho_{-,-\eta}\ + \ 
A_{-} \, \chi_{0} \, \rho_{-,1-\eta} \ + \ 
A_{+} \, \chi_{+} \, \rho_{-,-1-\eta} \quad .
\label{twchard4}
\ea
In terms of (\ref{untchard4}) and (\ref{twchard4}), it is relatively easy to
describe the perturbative spectrum of the open descendants.  The Klein bottle
is 
\be
{\cal{K}} \ = \ {1 \over 6} \, \Xi_{0,0} {\sum}^{'} \, e^{- \pi \tau
\alpha^{'} \, m_a G^{ab} m_{b} } \ + \ {1 \over 2} \ \Lambda_{0,0}
\qquad , \label{kleind4}
\ee
where the primed sum is over the non-zero modes.  
The form of ${\cal{K}}$ makes evident how 
in the closed GSO projection each character is paired with its conjugate,
but $\Lambda_{0,0}$ is the only self-conjugate one.  
We can thus anticipate that all
characters flow in the tube as well as in the
transverse Klein amplitude due to the presence of only $\Lambda_{0,0}$
in eq. (\ref{kleind4}).  Compatibility with the mirror-like involution
defining the annulus selects the sections corresponding to strings with
Neumann boundary conditions at both ends.  Only D-9-branes are thus present 
for this class of models and the annulus partition function takes the form
\ba
{\cal{A}} \ &=& \ {1 \over 6} \: (\, N \, + \, M \, + \, {\bar{M}}\,)^2  
\: \Xi_{0,0} \: {\sum}^{'}
\: e^{- \pi \tau \alpha^{'} \, m_a G^{ab} m_{b}} \ + \ {1 \over 2} \: (
N^2 + 2 M \bar{M} ) \: \Lambda_{0,0} \nonumber \\ &+& \ {1 \over 2} \: (
M^2 + 2 N \bar{M} ) \: \Lambda_{0,+} \ + \ {1 \over 2} \: ( {\bar{M}}^2
+ 2 N M ) \: \Lambda_{0,-} \quad .
\label{annulusd4}
\ea
The open spectrum is properly projected by the M\"obius-strip amplitude.  
A suitable basis of `hatted' real
characters can be defined, barring some subtleties, with the aid of the
rational model at $R_i = \sqrt{3}$, where an $SU(3)^{\otimes 3}$ symmetry is
present.  In terms of these characters, the result is
\be
{\cal{M}} \ = \ - \: {1 \over 6} \: Tr({\bf 1}) \: {\hat{\Xi}}_{0,0} \:
{\sum}^{'} \: e^{- \pi \tau \alpha^{'} \, m_a G^{ab} m_{b}} \ - \ {N
\over 2} \: {\hat{\Lambda}}_{0,0} \ - \  {M \over 2} \:
{\hat{\Lambda}}_{0,+} \ - \  {{\bar{M}} \over 2} \:
{\hat{\Lambda}}_{0,-} \quad, \label{moebiusd4}
\ee
where the negative signs anticipate the tadpole cancellations.  ${\cal{K}}$,
${\cal{A}}$ and ${\cal{M}}$ lead to sensible transverse channel amplitudes,
and the decoupling of unphysical states is ensured if
\ba
N \ + \ M \ + \ \bar{M} \ &=& \ 32 \nonumber\\
2 \: N \ - \ M \ - \ \bar{M} \ &=& \ - \: 8  \qquad ,
\label{tadpold4}
\ea
respectively from the untwisted and twisted massless sectors.  Because
of the (numerical) equality of $M$ and $\bar{M}$, eqs. (\ref{tadpold4}) are 
solved by  $N=8$ and $M=12$.

Let us analyze the resulting (closed and open) massless spectrum:  the
original type IIB closed spectrum is truncated to an $N=1$ supergravity
multiplet, coupled to the universal linear multiplet and to 9 chiral
multiplets from the untwisted sector, as well as to 27 chiral multiplets (one
for each fixed point) from the twisted sector.  The open unoriented spectrum
exhibits a Chan-Paton gauge group $SO(8) \otimes SU(12) \otimes U(1)$, with
three generations of chiral multiplets in the $({\bf{8}},{\bf
{12}}^{\ast})_{-1}$ and in the $({\bf{1}},{\bf{66}})_{2}$.  Absence of
anomalies is guaranteed by tadpole cancellations, apart from the $U(1)$
factor, which is disposed of by a Higgs mechanism giving a large mass to
the corresponding $U(1)$ gauge boson.  This is the analogue of what happens in
the heterotic theory \cite{dsw}, 
and in fact the relation between heterotic and type I
dilaton in $d$ dimensions is \cite{chiral}
\be
\phi^{(d)}_I \ = \ {6-d \over 4} \ \phi^{(d)}_H \ - 
\ {{(d-2)} \over {16}} \log \det G^{(10-d)}_H \quad ,
\label{dualityIH}
\ee   
where $G^{(10-d)}_H$ is the internal metric in the  heterotic-string
frame.  The duality between Type I and Heterotic models
in $d=10$ survives in $d=4$, but becomes a weak-weak coupling duality 
rather than a strong-weak one.  In
\cite{chiral} a candidate heterotic dual to this chiral type I model is
described, with additional chiral multiplets in the $({\bf 8_c}, {\bf 1})$ 
from twisted sectors.  
A final remark is concerned with the K\"ahler manifold
of the untwisted scalars in the low-energy supergravity \cite{wser}.  This
must be a K\"ahler submanifold of the corresponding quaternionic manifold of
the parent type IIB orbifold.  The unoriented truncation, however, is highly
non-trivial due to several mixtures between moduli to give the `real
slice'.  The manifold can be uniquely selected to be $Sp(8,R)/(SU(4) \times
U(1))$.  The $SU(4)$ factor is suggestive of the relation between these 
type I vacua and the putative (12-d) F-theory compactified on
Calabi-Yau fourfolds.

To summarize, we have analyzed some instances of perturbative type I vacua
stressing the elegance of the construction and some `exotic' features
emerging from it.  The interest in the subject is increasing, in
light of some recent results suggesting that D-branes and associated 
open-string excitations play a role as 
fundamental degrees of freedom in
the `complete' non-perturbative formulation of superstring theories 
\cite{dbrane}.

\vskip 24pt \begin{flushleft}
{\large \bf Acknowledgments}
\end{flushleft}

I would like to thank the organizers for the kind invitation and 
C. Angelantonj, M. Bianchi, S. Ferrara, A. Sagnotti and Ya.S.
Stanev for the stimulating collaboration.  This work was supported in part by 
ECC Grant CHRX-CT93-0340.

\end{document}